\documentclass[reprint,aps,prl,superscriptaddress,amsmath,amssymb,showpacs,preprintnumbers,twoclumn]{revtex4-1}
\usepackage{amsmath}
\usepackage{amssymb}
\usepackage{amsfonts}
\usepackage{dcolumn}
\usepackage{bm}
\usepackage{times}
\usepackage{comment}
\usepackage{latexsym}
\usepackage{subfigure}
\usepackage{hyperref}
\usepackage{graphicx}
\usepackage{upgreek}
\usepackage{color}
\begin{document}
\bibliographystyle{unsrt}
\newcommand{\RNum}[1]{\uppercase\expandafter{\romannumeral #1\relax}}
\title{Entanglement Distribution in Star Network Based on Spin Chain in Diamond}
\author{Yuan-Ming Zhu}
\author{Lei Ma}
\affiliation{College of Physics and Materials Science, East China Normal University,
Shanghai 200241, China\\ \vspace{0.2cm}}
\date{\today\\ \vspace{0.2cm}}
\email[Email: ]{lma@phy.ecnu.edu.cn}

\begin{abstract}
Since star network of spins was proposed, generating entanglement directly through spin interactions between distant parties became much possible. We propose an architecture which involves coupled spin chains based on nitrogen-vacancy centers and nitrogen defect spins to expand star network, the numerical analysis shows that the length of spin chains $M$ and spin noise can determine the maximally achievable entanglement $E_m$. The entanglement capability of this configuration under effect of disorder and spin loss is also studied. Moreover, it is shown that with this kind of architecture, star network of spins is feasible in measurement of magnetic-field gradient.

\end{abstract}
\maketitle
\section{Introduction.}
Quantum state transfer and entanglement distribution are necessary in a solid quantum information processing system\cite{PhysRevLett.91.207901, PhysRevA.71.032310, PhysRevA.71.052315}. However, most short distance interaction strength decays rapidly on account of physical separation. Thus it is significant to have appropriate physical system which can serve as channel for quantum communication. In star network configuration\cite{PhysRevA.69.042312}, one can create entangled state easily according to the number of spins. What restricts the application of this configuration is that the entangled particles will be created very close to each other, usually at the nanoscale. If interactions between qubits can be made long range, then this structure could be used for entanglement distributions between several distant parties.

The nitrogen-vacancy (NV$^-$) centers in diamond are excellent quantum processor in solids\cite{Fuchs1520, DOHERTY20131}. The spin qubits of NV$^-$ center can be optically initialized and read out\cite{Gruber2012} and has a long spin-coherence time up to milliseconds even at room temperature\cite{Maurer1283, Gaebel}. Meanwhile, many quantum gates\cite{PhysRevLett.93.130501, PhysRevLett.112.050503}, algorithms\cite{PhysRevLett.105.040504, PhysRevLett.118.130504} and simulations\cite{PhysRevLett.117.060503} have been demonstrated on NV scheme as well. Recently, NV$^-$ center has been used to detect AC magnetic fields\cite{PhysRevLett.106.080802}, magnetic noises\cite{PhysRevLett.103.220802} and single nuclear spins\cite{PhysRevLett.109.137602} via dynamical decoupling. All of above features may be used for extending the star configuration of spins to practical range.

In recent years, a proposal suggested a chain which consists of implanted nitrogen impurities as a coherent quantum channel to transfer quantum states between distant nitrogen-vacancy centers at room temperature\cite{PhysRevLett.106.040505}. The NV$^-$ centers interact with nitrogen defects through dipole-dipole coupling. This protocol alleviated the stringent constraints currently limiting the realization of scalable quantum processors. The later research\cite{PhysRevLett.110.100503} extended nearest-neighbor dipole-dipole coupling to non-nearest-neighbor coupling, proved that distribution of a finite amount of entanglement appears realistic with current systems.

In this paper, we describe and numerically simulate a feasible star configuration of spins for entanglement distribution, the practical application of this configuration in measurement of magnetic-field gradient is also discussed. This work is inspired by scheme of star network of spins and the rise of research on NV technologies. More details are described in Sec.\RNum{2} \& \RNum{3}.

\section{Physical model and methods.}
\begin{figure}
\includegraphics[width=0.8\columnwidth]{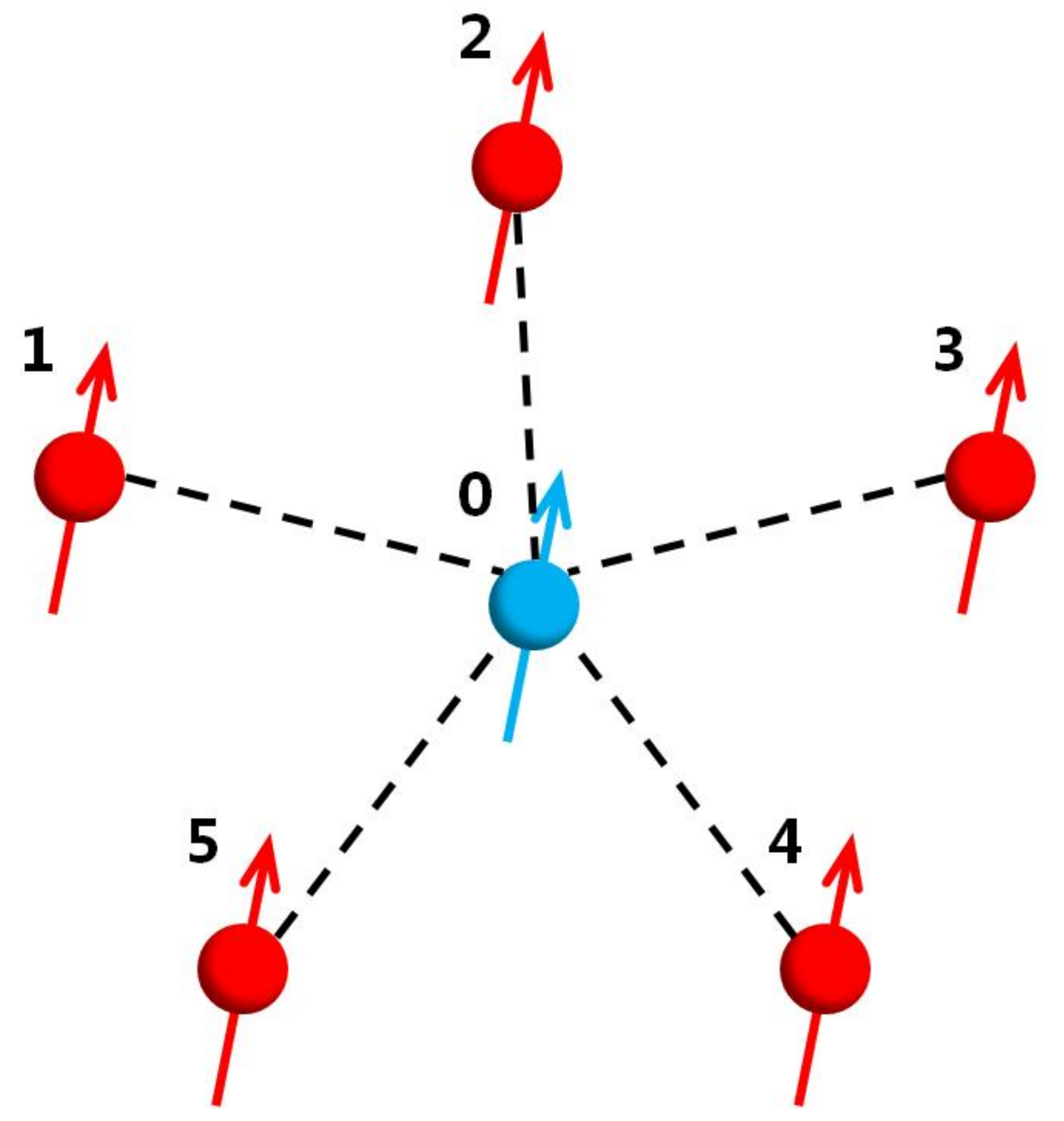}
\caption{\ (color online). Star configuration of spins, central spin is labeled 0 and spins labeled 1-5 interact with central spin.}
\label{fig1}
\end{figure}

First we give an overview of the star network of spins\cite{PhysRevA.69.042312}, The structure is depicted in Fig. \ref{fig1}, where 0 depicts the central spin. The spins 1-5 interact only with the central spin and not with each other. The ground state of this configuration is an interesting multiparticle entangled state, symmetric in the outer spins. The key of the entanglement between the outer spins at zero temperature to a degree is the total number of spins in the spin star. The Hamiltonian of this system is given by
\begin{equation}
H=\lambda(\sigma_{0x}\sum_{\rm outer}\sigma_{ix}+\sigma_{0y}\sum_{\rm outer}\sigma_{iy}).
\end{equation}
where the summation over ¡°outer¡± refers to the outer spins, $\lambda$ is coupling strength between central spin and outer spins, $\sigma_{ix}$ and $\sigma_{iy}$ denote the $\sigma_{x}$ and $\sigma_{y}$ Pauli operators for the $i$th outer spin, $\sigma_{0x}$ and $\sigma_{0y}$ denote the Pauli operators for the central spin. It is convenient to use $J_x=\sum_{\rm outer}\sigma_{ix}$, $J_y=\sum_{\rm outer}\sigma_{iy}$ and $J_z=\sum_{\rm outer}\sigma_{iz}$ where $J=iJ_x+jJ_y+kJ_z$ ($\hbar=1$). Accordingly, we can therefore rewrite the Hamiltonian in Eq. (1) with the raising and lowering operators $\sigma_{\pm}=(\sigma_{x} \pm i\sigma_{y})/2$ and $J_{\pm}=\sum_{\rm outer}{\sigma_{\pm}}$ as
\begin{equation}
H=\lambda(\sigma_{0+}J_{-}+\sigma_{0-}J_{+}).
\end{equation}
This Hamiltonian can be seemed as a resonant interaction between a spin 1/2 and a higher spin scheme. It is readily seen that such a scheme is similar to Jaynes-Cummings Model, especially in the form of Hamiltonian. Thus the eigenstates of this Hamiltonian yield the form
\begin{equation}
\frac{1}{\sqrt{2}}(|0\rangle|{j,m}\rangle\pm |1\rangle|{j,m-1}\rangle).
\end{equation}
where the states $|0\rangle$ and $|1\rangle$ represent the $|-1/2\rangle$ and $|1/2\rangle$ spin states of central spin, and the second ket is an eigenstate of $J^2$. Both $m$ and $m-1$ range from $-j$ to $j$, therefore $-j+1\leq m \leq j$. The energy eigenvalues for the star-spin system are given by
\begin{equation}
E=\pm\lambda\sqrt{(j+m)(j-m+1)}.
\end{equation}
To focus on the ground state, we assume $\lambda$ positive and the number of outer spins be \emph{N}. For the case \emph{N} odd, the energy eigenvalue will be lowest when $m=1/2$, i.e., the eigenstate
\begin{equation}
|\psi_{\rm odd}\rangle=\frac{1}{\sqrt{2}}(|0\rangle|{N/2,1/2}\rangle-|1\rangle|{N/2,-1/2}\rangle).
\end{equation}
and if \emph{N} is even, the ground state is degenerate because there are two states with the lowest possible energy, when $m=0$ or $m=1$:
\begin{equation}
|\psi_{\rm even1}\rangle=\frac{1}{\sqrt{2}}(|0\rangle|{N/2,0}\rangle-|1\rangle|{N/2,-1}\rangle).
\end{equation}
\begin{equation}
|\psi_{\rm even2}\rangle=\frac{1}{\sqrt{2}}(|0\rangle|{N/2,1}\rangle-|1\rangle|{N/2,0}\rangle).
\end{equation}
For example, for \emph{N} = 3, we can get these familiar \emph{W}$_3$ states
\begin{equation}
|3/2,1/2\rangle=\frac{1}{\sqrt{3}}(|011\rangle+|101\rangle+|110\rangle).
\end{equation}
\begin{equation}
|3/2,-1/2\rangle=\frac{1}{\sqrt{3}}(|001\rangle+|010\rangle+|100\rangle).
\end{equation}
To conclude this brief introduction to this scheme, one can achieve a \emph{N}-qubit \emph{W}-states through measurement on central spin, and with measuring the outer spins, a 2-qubit maximally entangled state will be created as well.

With the rising interest on NV technologies, we can also utilize NV$^-$ center as individual spins. However, the dipole-dipole interaction between NV$^-$ centers is too weak (the coupling strength is about 26kHz when separation between spins is 10 nm), therefore entangled particles would be created very close to each other. In this case, NV$^-$ center qubits would lose individual addressability\cite{Maurer}. To assure the star configuration work, spin chain structure\cite{PhysRevLett.91.207901} will be useful to extend the separation between spins. Nitrogen defects consist media channel in spin chain.

\begin{figure}
\includegraphics[width=0.9\columnwidth]{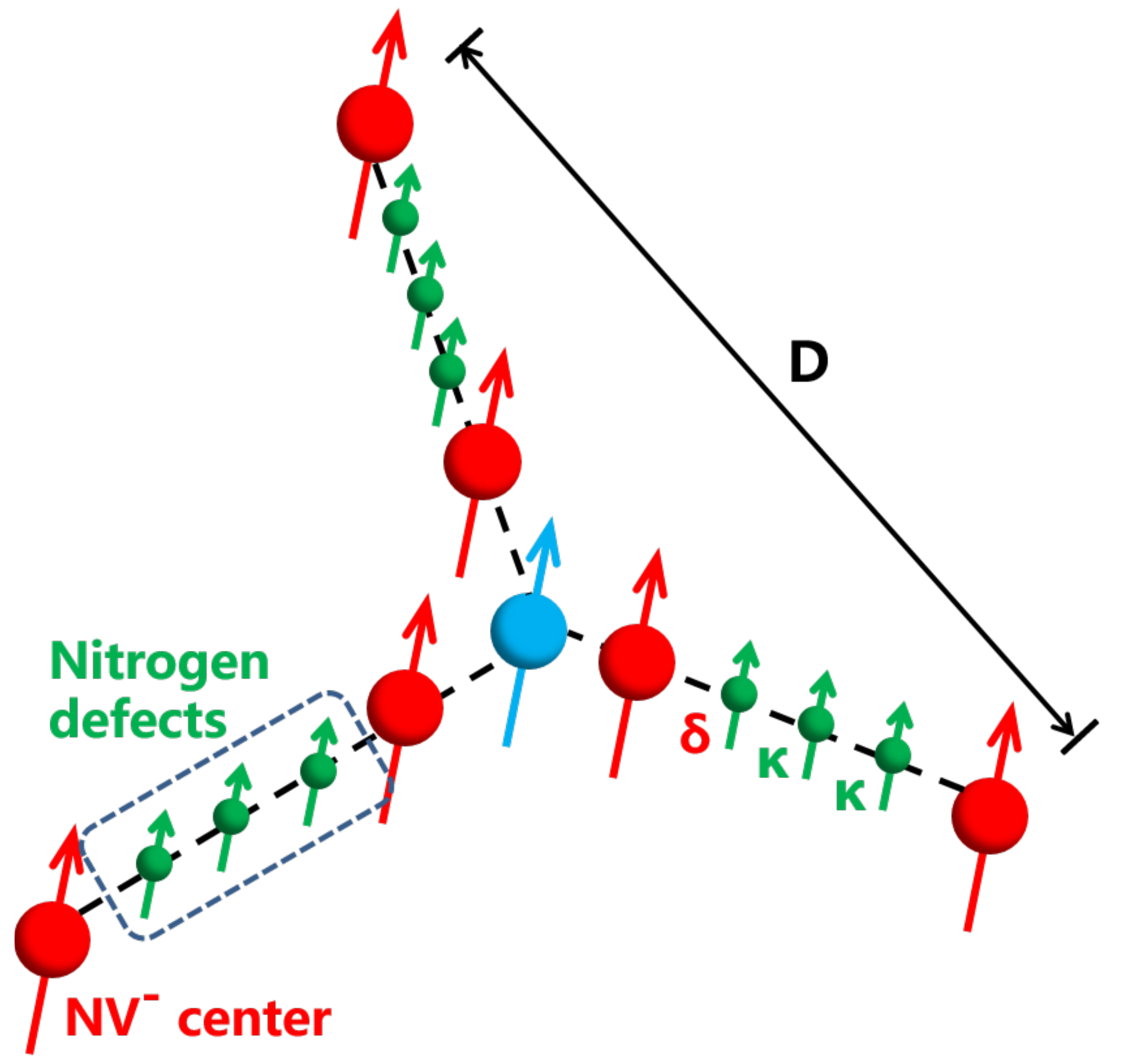}
\caption{\ (color online). Conceptual schematic of our structure. Star configuration is expanded by spin chains consisting of $N$ nitrogen defects, all spins in chain interact with each others through nearest-neighbor and second-nearest-neighbor dipole-dipole coupling. $\kappa$ and $\delta$ are the dipolar coupling strength between Nitrogen-Nitrogen and Nitrogen-NV$^-$ center, respectively. The separation between outer NV$^-$ centers is $D$.}
\label{fig2}
\end{figure}

From Ref.\cite{PhysRevLett.110.100503}, we derive the following effective Hamiltonian for a chain consisting of $M$ spins plus two register spins located at either end of the chain (Fig. \ref{fig2})
\begin{equation}
H_{\rm eff}=\sum_{i=1}^{M-1}\kappa\sigma_{+}^{i}\sigma_{-}^{i-1}+\sum_{j=0,M}\delta\sigma_{+}^{j}\sigma_{-}^{j-1}+\rm{H.c.}.
\label{10}
\end{equation}
where $\sigma_{\pm}^{i}=\sigma_{x}^{i}\pm i\sigma_{y}^{i}$ denote Pauli operators acting on \emph{i}th spin, the coupling strengths $\kappa$ and $\delta$ (actually in our simulation, $\delta$ can be tuned to $\delta=0.9\kappa$ for better performance  Ref.\cite{PhysRevLett.110.100503}) are proportional to distance $1/r^3$. H.c. represents the Hermitian conjugate. We simulate the full spin chain dynamics by numerically integrating a Lindblad master equation using Hamiltonian in Eq. (\ref{10})
\begin{equation}
\dot{\rho}=-i[H_{\rm eff},\rho]+\sum_{i=0}^{M+1}\Gamma_{i}(L_{i}\rho L_{i}^{\dag}-1/2(L_{i}^{\dag}L_{i}\rho+\rho L_{i}^{\dag}L_{i})).
\end{equation}
where $\Gamma_i$ and $L_{i}$ denote the noise rates and Lindblad operators, respectively. Considering that phase coherence time ($T_2$) of NV$^-$ center or nitrogen defect is far less than energy relaxation time ($T_1$), we only discuss $T_2$-like process, all noise rates are given by $\Gamma_i=1/{T_2}$, and $L_{i}=\sigma_{z}^{i}$.

Before further discussion, there is a restriction on the total number of outer spins. Because both nearest-neighbor and second-nearest-neighbor interactions are included, and $\delta/\kappa$ is fixed to 0.9, the distance $D$ between outer spins should be greater than 2$r$, hence \emph{N} and \emph{M} yield the inequality
\begin{equation}
\rm \sin(\frac{\pi}{2\emph{N}})>\frac{1}{\emph{M}+\frac{20\sqrt{10}}{3}}.
\label{12}
\end{equation}
\begin{figure}
\includegraphics[width=0.9\columnwidth]{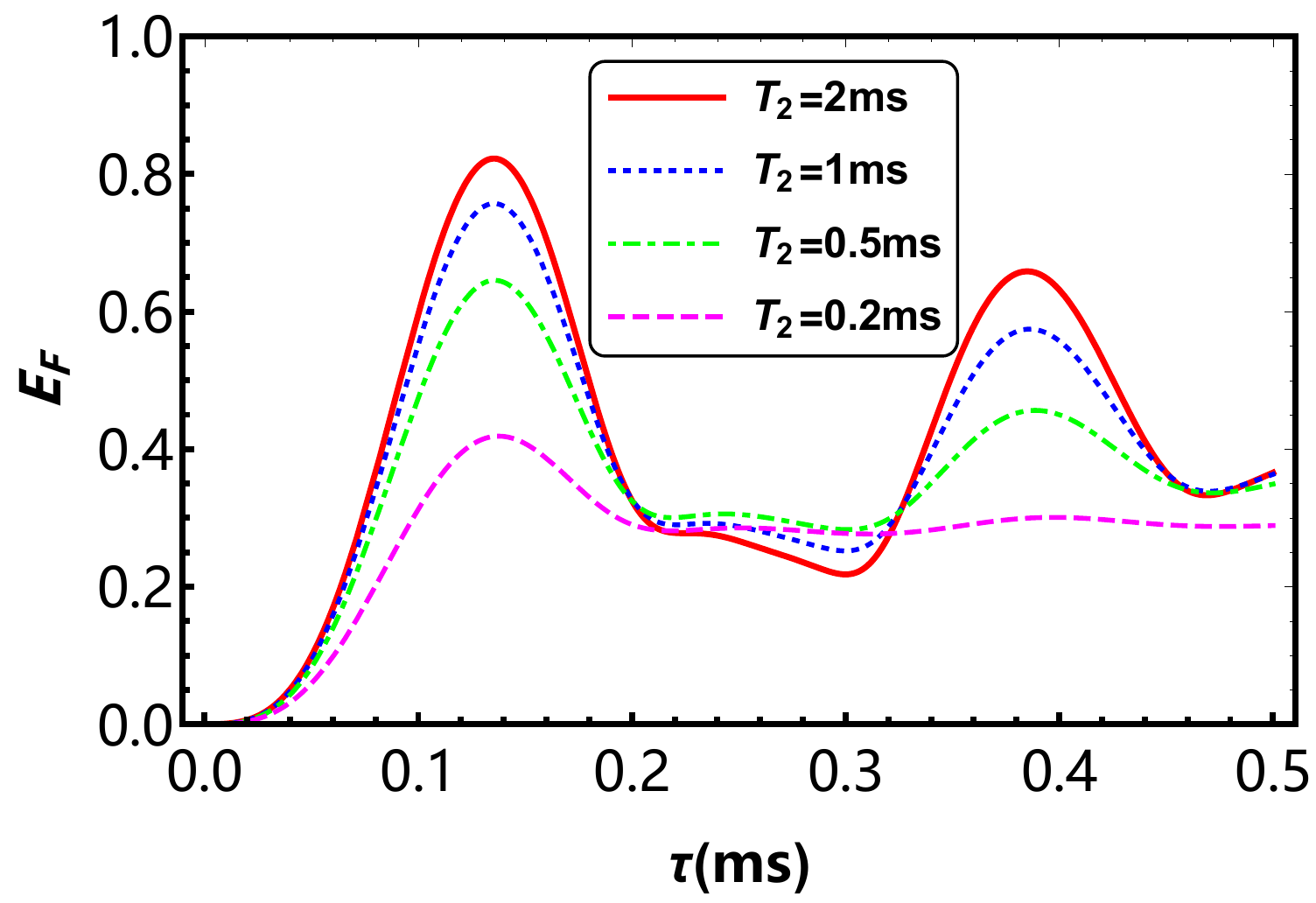}
\caption{\ (color online). Entanglement of formation $E_F$ between two outer spins versus transfer time $\tau$ with noise rate $\Gamma=1/{T_2}$, the length and initial state of both branch of spin star are \emph{M} = 3 and $|+\rangle|000\rangle|0\rangle$, respectively. $\delta/\kappa$ is fixed to 0.9 for $\kappa$ = 26 kHz ($r$ = 10 nm).}
\label{fig3}
\end{figure}

Now we were supposed to test the feature of star configuration, which is, the entanglement of \emph{W}-states. Unfortunately we haven't found an appropriate way to measure the entanglement of mixed \emph{W}-states\cite{PhysRevA.61.052306} directly, thus we have to keep measuring the outer spins to create a 2-qubit maximally entangled state. To reduce the complexity of operation, the number of outer spins is \emph{N} = 3, i.e., this scheme can create \emph{W}$_3$ states.

\section{Results.}
\begin{figure}
  \centering
  \subfigure[]{\label{fig4a}\includegraphics[width=0.9\columnwidth]{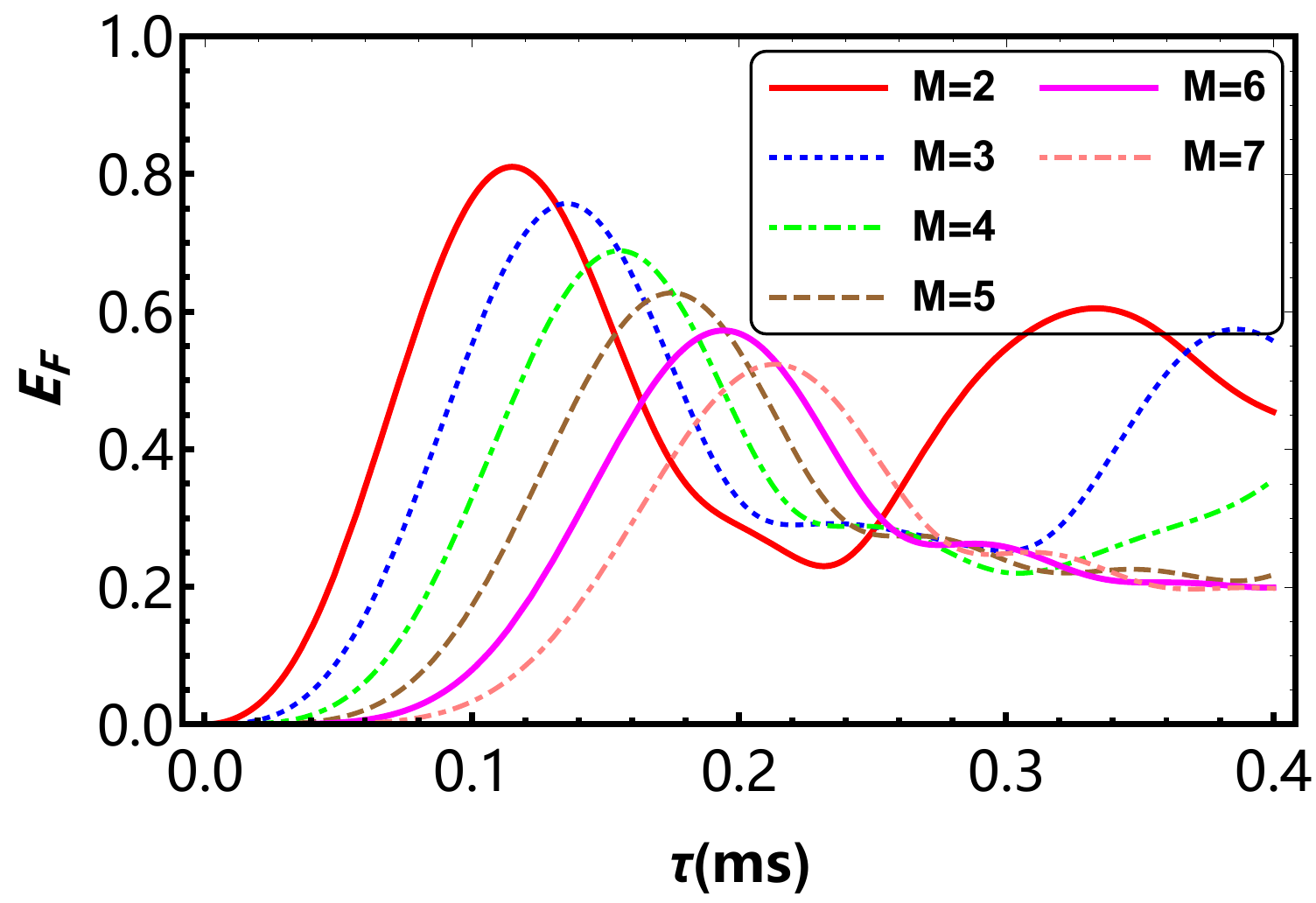}}
  \subfigure[]{\label{fig4b}\includegraphics[width=0.9\columnwidth]{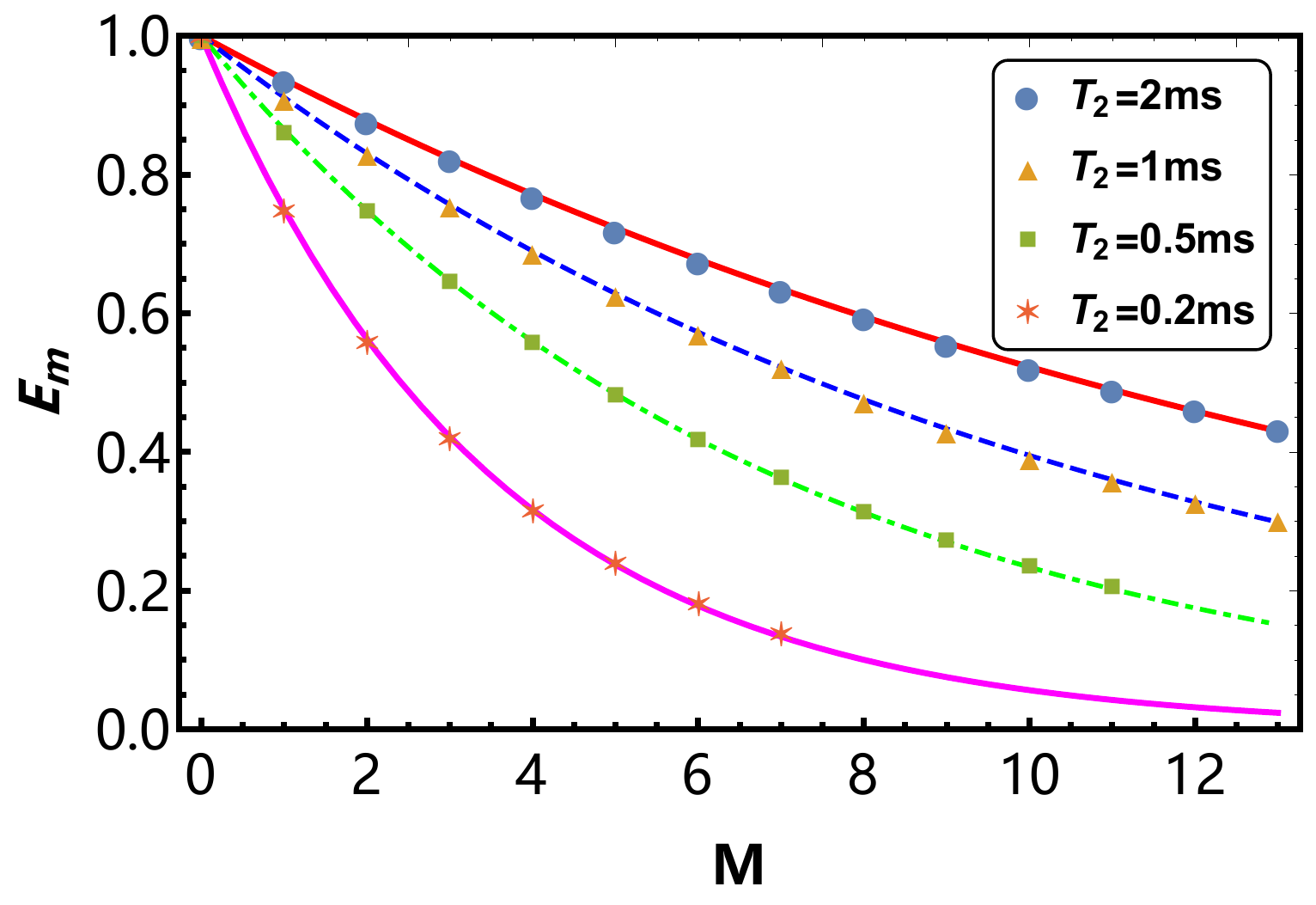}}
  \caption{\ (color online). (a) Entanglement of formation $E_F$ versus transfer time $\tau$ in longer chains with $T_2$ = 1 ms. (b)Maximally achievable entanglement $E_m$ between outer spins versus length of chain.}
  \label{fig4}
\end{figure}

Let us omit all noise created in measurement process, consider both nearest-neighbor and second-nearest-neighbor interactions. Starting with \emph{M} = 3 as the shortest nontrivial odd chain (notice that value of \emph{M} and \emph{N} satisfies Eq. (\ref{12})). Fig. \ref{fig3} shows that $T_2$ has damaging effect on the chain. However, unlike the fidelity of quantum state or quantum gate in practical, it is not necessary for entanglement of formation to reach at 0.9 or higher level, thus this property allow us to study longer chains(Fig. \ref{fig4a}). Unsurprisingly, the performance of the chain is becoming worse drastically as long as \emph{M} getting larger, because more spins are exposed to the noise. We also calculate the maximally achievable entanglement $E_m$ (Fig. \ref{fig4b}), on account of deficiency of data (the demand for memory of computation is exponential), we may not derive precise expression, but still can find the $E_m$ exhibit exponential dependence on $T_2$ and \emph{M}:
\begin{equation}
E_m\propto \exp[{-a(\frac{1}{T_{2}})^{b}M}].
\end{equation}
where parameter a,b depends on the value of $\delta/\kappa$ and initial state of chain. Simulation results show feasibility of distributing entanglement in outer spins.

\subsection{The Effect of Coupling-Strength Disorder.}
\begin{figure}[!h]
\includegraphics[width=0.9\columnwidth]{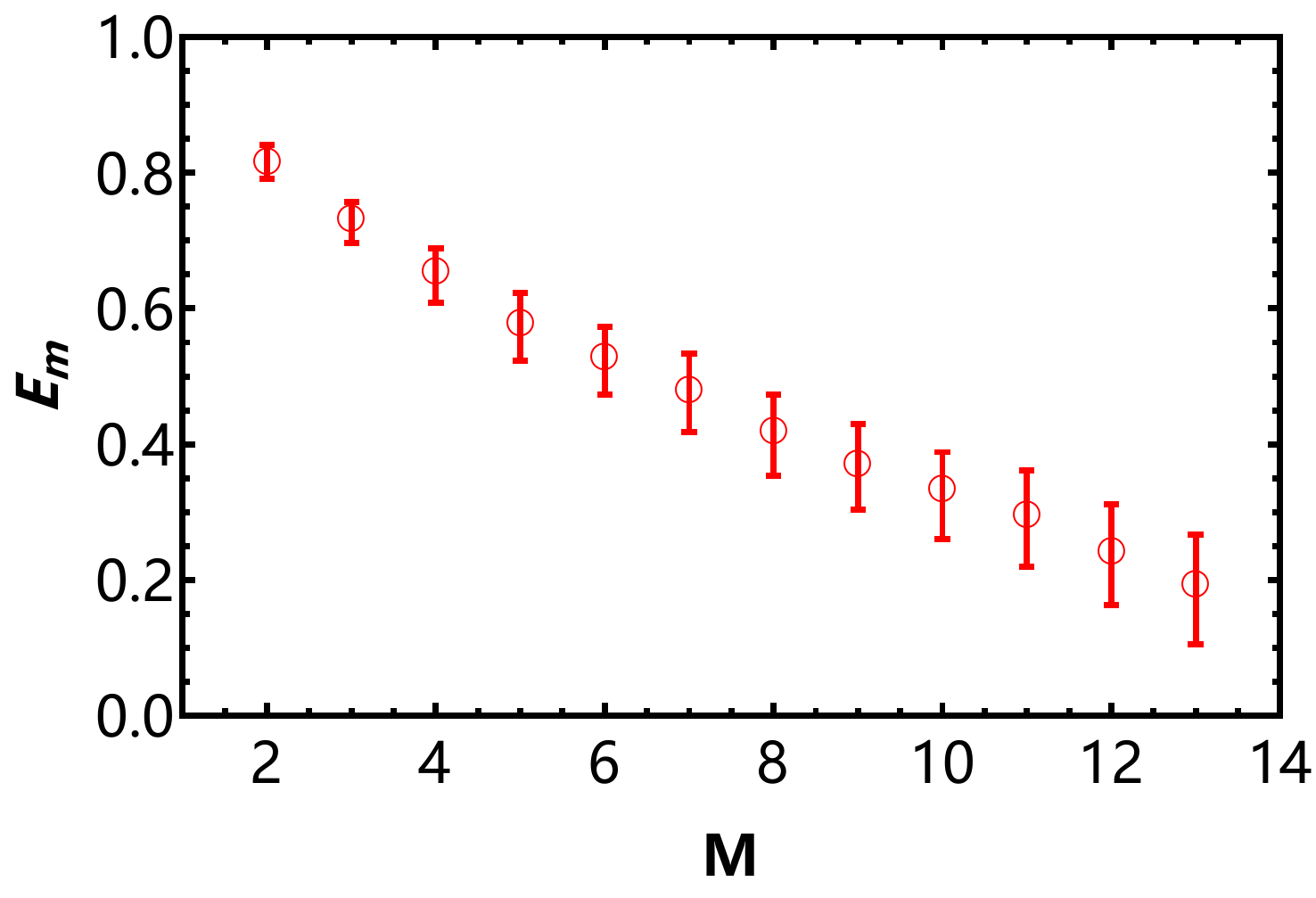}
\caption{\ (color online). Maximally achievable entanglement $E_m$ versus length \emph{M} under effect of disorder, the spacings between neighbouring spins obey a Normal distribution around the mean value $r$ = 10 nm with variance $\sigma^2=0.25$, $T_2$ = 1 ms, each data point consists of one hundred independent runs. }
\label{fig6}
\end{figure}

In experiment, the imperfection of implantation will cause disorder, coupling strength usually is affected by the inhomogeneous of space interval ($r$) in chain. We assume the spacings between neighbouring spins obey a Normal distribution around the mean value $r$ = 10 nm, and choose 5\% disorder on $r$. Our each data point consists of one hundred independent runs. Fig. \ref{fig6} shows that 5\% disorder on $r$ does not have severe impact on $E_m$ when length is not quite long (\emph{M} $<$ 11). But as the chain becomes much longer, effect of disorder will cause more chaos in chain.
\subsection{The Effect of Spin Loss.}
\begin{figure}[!h]
  \centering
  \subfigure[]{\label{fig7a}\includegraphics[width=0.48\columnwidth]{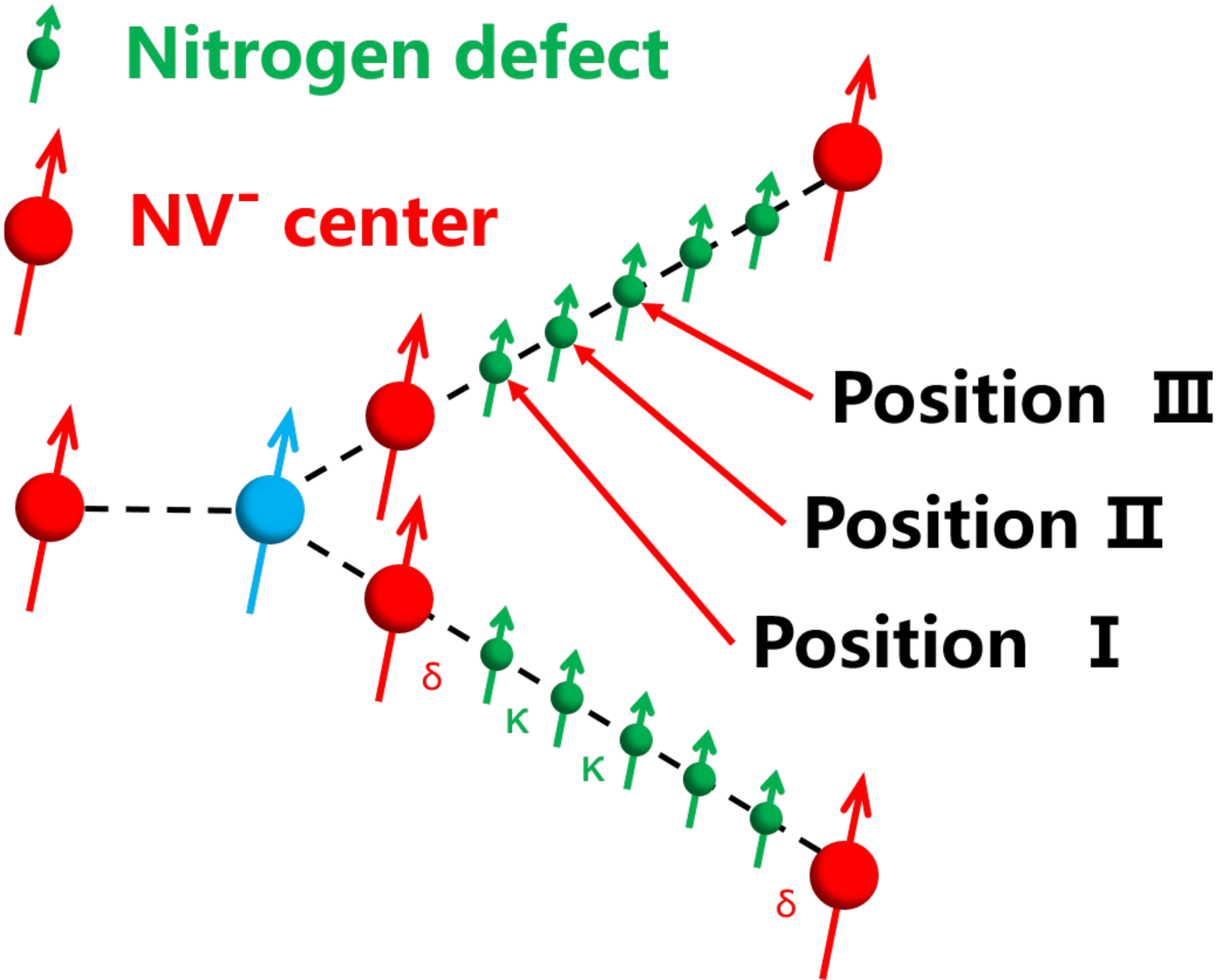}}
  \subfigure[]{\label{fig7b}\includegraphics[width=0.51\columnwidth]{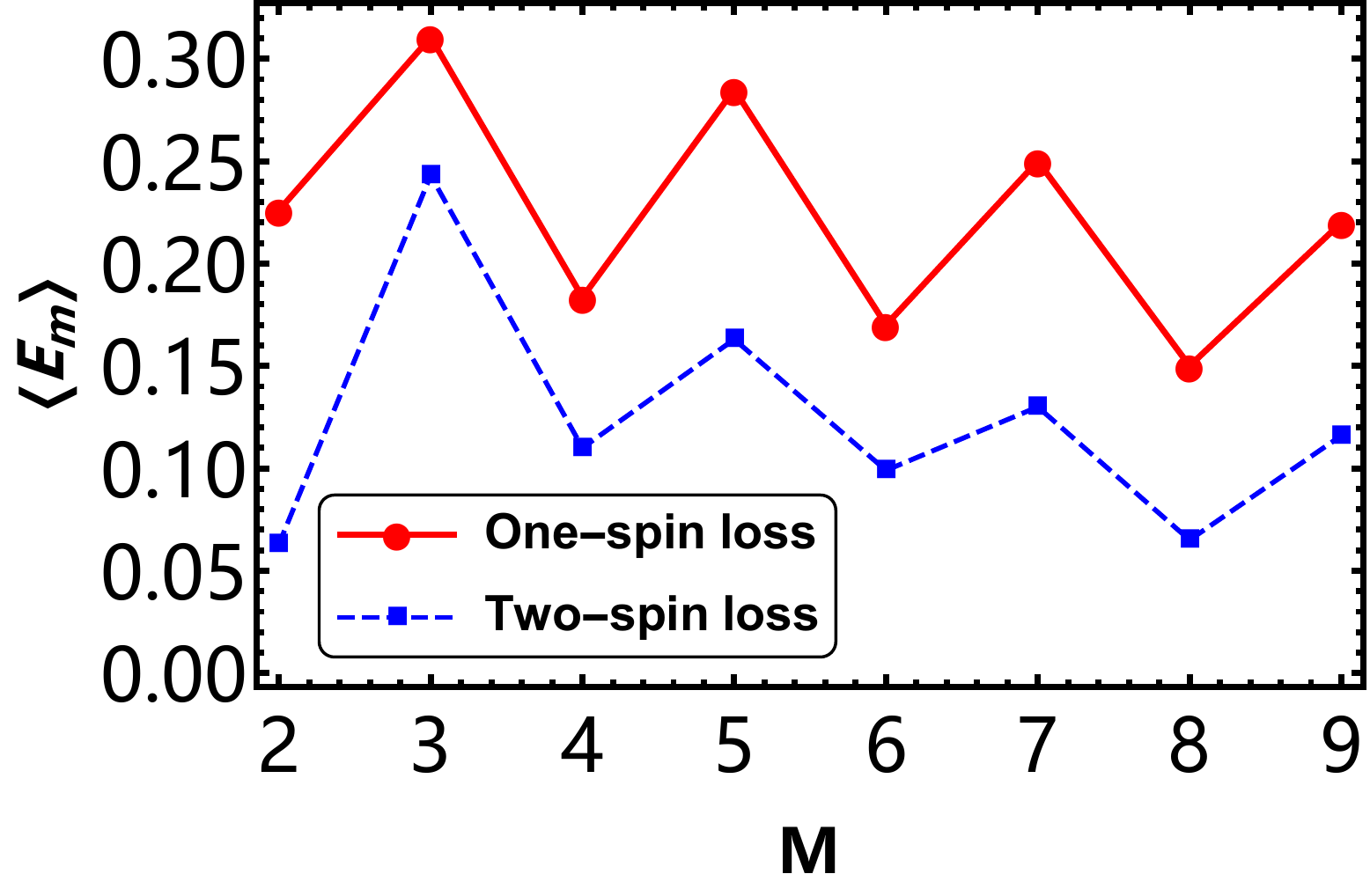}}
  \subfigure[]{\label{fig7c}\includegraphics[width=0.9\columnwidth]{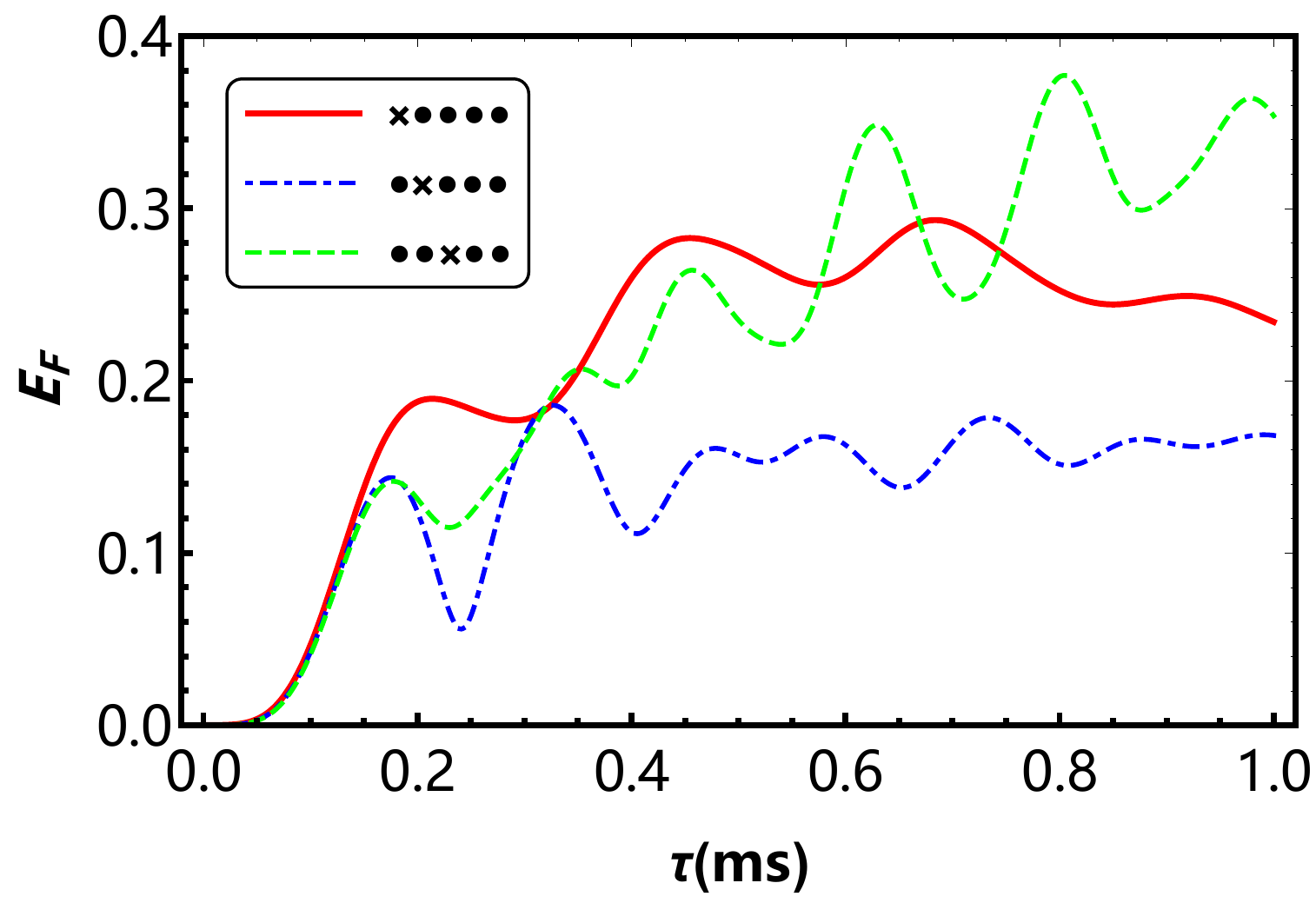}}
  \subfigure[]{\label{fig7d}\includegraphics[width=0.9\columnwidth]{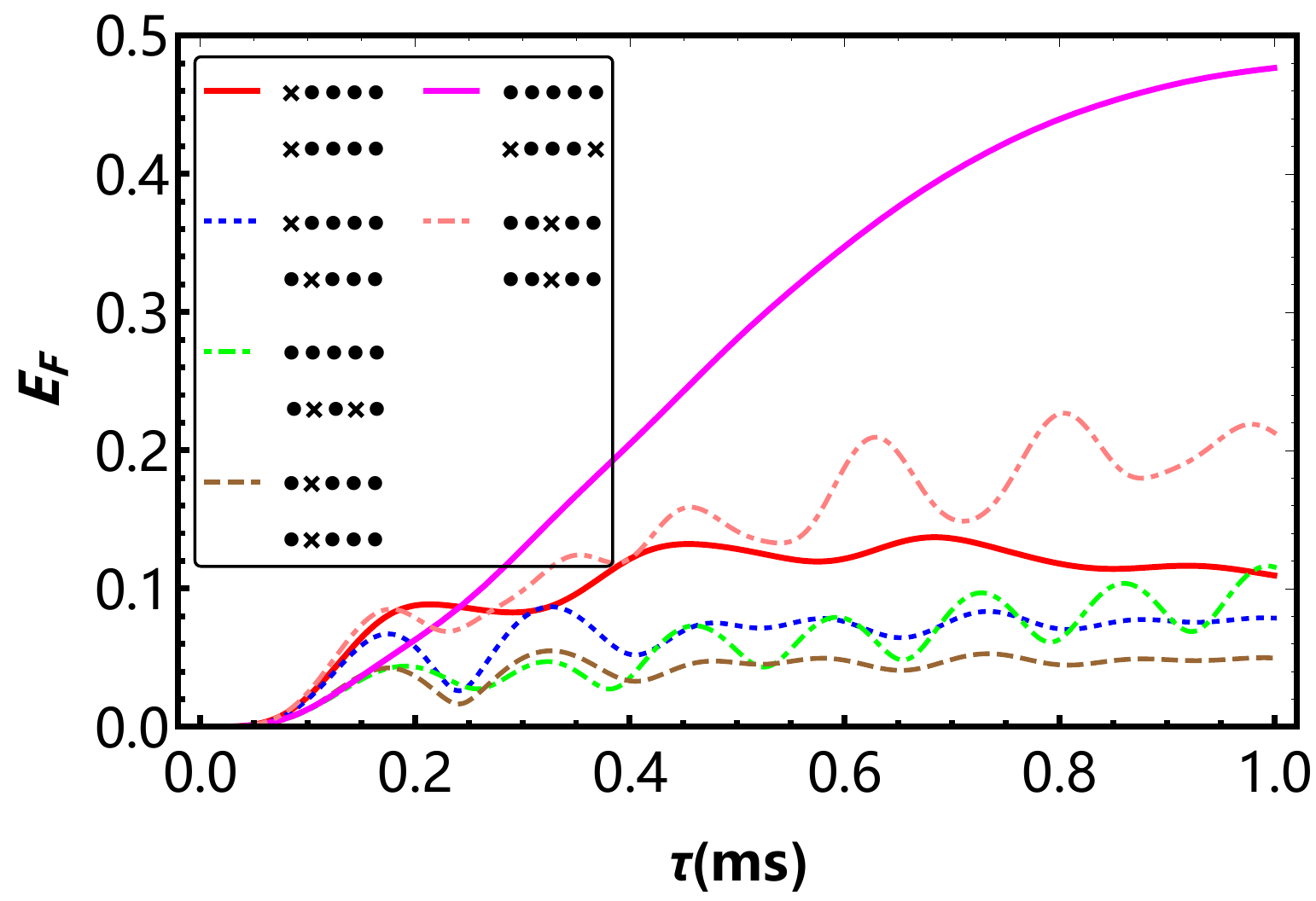}}
  \caption{\ (color online). (a)Schematic diagram of spin loss in position \RNum{1}, \RNum{2} \& \RNum{3}. (b)Expectation of $E_m$ versus length \emph{M} with one-spin loss and two-spin loss progress. Entanglement of formation $E_F$ versus transfer time $\tau$ in \emph{M} = 5 chains under effect of one-spin loss(c) and effect of two-spin loss(d), the dots in linelegend denote spins while the crosses represent spin losses, $T_2$ = 1 ms.}
  \label{fig7}
\end{figure}
Now we study the entanglement distribution in the presence of spin loss. Assuming that only one-spin loss and two-spin loss behaviors exist, we calculate $E_m$ at all possible place that spin loss will occur. Because we only consider nearest-neighbor and second-nearest-neighbor dipole-dipole coupling, loss of two neighboring spins (which means this protocol is hardly working) are not included in our simulation. Because of the symmetry in \emph{M} = 5 chain (Fig. \ref{fig7a}), we only focus on spin loss occur in position \RNum{1}, \RNum{2} \& \RNum{3}, Fig. \ref{fig7c} and Fig. \ref{fig7d} show that spin loss in different position will cause different dynamics of $E_F$.

In some special cases, two-spin loss can achieve even better result than one-spin loss effect. For example, the chain which lost first and last spin simultaneously will achieve the highest $E_m$ in Fig. \ref{fig7d}. Because that chain now can be considered as a \emph{M} = 3 chain with $\delta/\kappa$ = 0.12. Fig. \ref{fig7b} denotes expectation of $E_m$ versus length \emph{M} with one-spin loss and two-spin loss progress, the value of $<E_m>$ are calculated by averaging over all possible missing spin configurations in two branches. It is clearly that there are oscillations with even and odd \emph{M} in the creation of a expectation of $E_m$ between two outer spins. Spin chain with odd number is more robust against effect of spin loss than those with even number, this result is in agreement with Ref.\cite{PhysRevLett.106.040505, PhysRevLett.110.100503}.

\section{Detection of magnetic-field gradient.}
In addition to quantum information processing, spins star may be used in other area, we investigate the feasibility of measurement of magnetic-field gradient with star network of spins. First we consider an atom to have two hyperfine spin states. By measuring the difference of the transition frequencies of two atoms at the two different locations, the magnetic-field gradient can be determined. Based on above assumption, several methods\cite{PhysRevA.87.043602, NG2014353, PhysRevA.88.013626} have been proposed, and using star configuration can both detect magnitude and direction of magnetic-field gradient easily as well.

\begin{figure}
\centering
  \subfigure[]{\label{fig8a}\includegraphics[width=0.37\columnwidth]{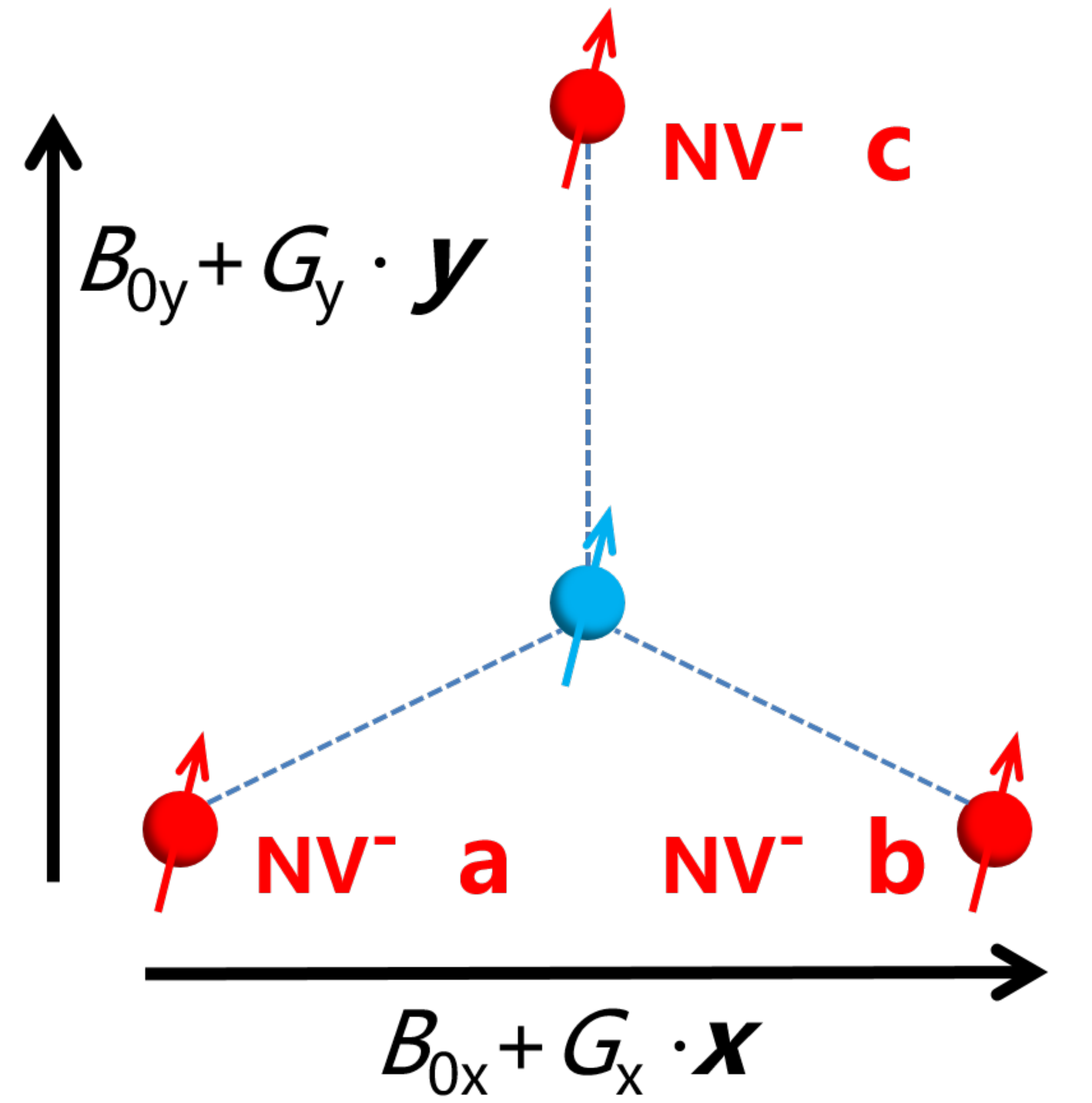}}
  \subfigure[]{\label{fig8b}\includegraphics[width=0.62\columnwidth]{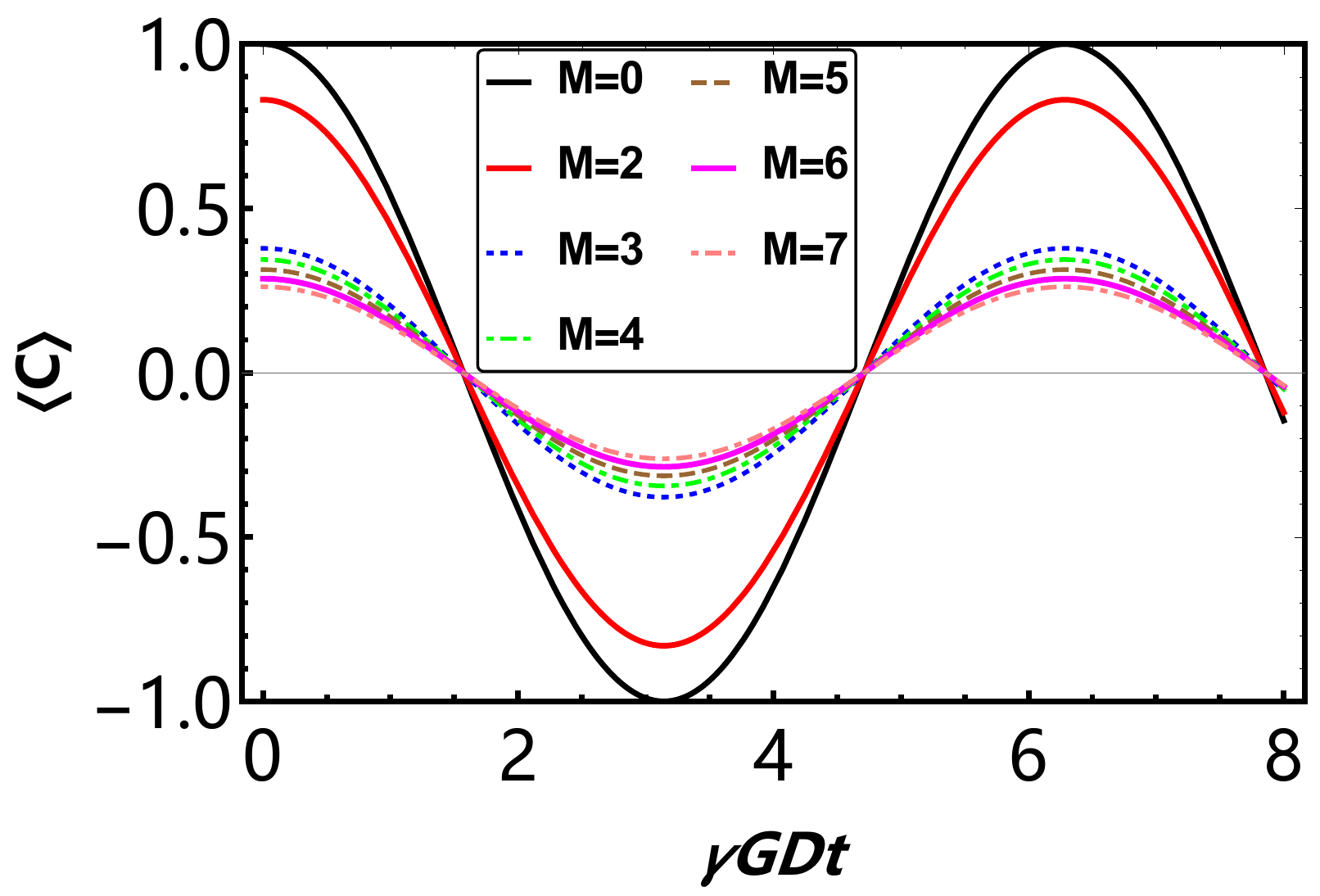}}
\caption{\ (color online). (a)Schematic diagram of detection of magnetic-field gradient with star configuration. (b)Quantum coherence factor $\langle C\rangle$ versus $\gamma GDt$ for different length \emph{M}, $T_2$ = 1 ms.}
\label{fig8}
\end{figure}

For reducing complexity, we still choose \emph{N} = 3. The assumptive magnetic field (Fig. \ref{fig8a}) linearly varies with the position
\begin{equation}
B(\emph{\textbf{x}},\emph{\textbf{y})}=B_{0}+G_{_{x}}\emph{\textbf{x}}+G_{_{y}}\emph{\textbf{y}}.
\end{equation}
where $B_{0}$ is the reference magnetic field, $G_{_{x}}$ and $G_{_{y}}$ is the magnetic-field gradient along X and Y axis, respectively. The transition frequency $\omega$ rely on the magnetic-field gradient $\omega=\omega_{0}+\gamma B(x,y)$, where $\gamma$ denotes gyromagnetic ratio and $\omega_{0} $ is transition frequency without the external magnetic-field. The 2-qubit maximally entangled states of NV$^-$ centers $\psi$ becomes $e^{-2i\omega t}|10\rangle +e^{-4i\omega t}|01\rangle$ under gradient magnetic-field,
Consider that \emph{C} is quantum coherence factor
\begin{equation}
\emph{C}=2\rho-(\rho_{_{10}}+\rho_{_{01}}).
\end{equation}
where $\rho$ denotes the density matrix of the entangled NV$^-$ center spins, $\rho_{_{10}}=|10\rangle\langle10|$ and $\rho_{_{01}}=|01\rangle\langle01|$. Omitting global phase factor, the expectation value of \emph{C} is given by
\begin{equation}
\langle C\rangle=\langle \psi|C|\psi\rangle=\cos(\gamma GDt).
\end{equation}
$\langle C\rangle$ is function of gradient $G$ and can easily be experimentally measured\cite{NG2014353}, i.e., after first time measurement (making NV$^-_a$ and NV$^-_b$ entangled) on the quantity of $\langle C\rangle$, we can derive gradient $G_{_{x}}$, then the second round trial (making NV$^-_a$ and NV$^-_c$ entangled) will figure out the value of $G_{_{y}}$. Fig. \ref{fig8b} shows coherence factor $\langle C\rangle$ versus $\gamma GDt$ for different length \emph{M}, as \emph{M} getting larger, the precision of measurement is much lower. Thus when NV$^-$ centers are distant enough to be addressable, it is better for this protocol to use shorter chain in detection.

\section{Conclusion.}
In conclusion, through numerical simulation we construct a feasible spin chain structure to extend star network. This configuration could be used for creating \emph{N}-qubit \emph{W}-states and entanglement distribution between several distant parties after extension. The performance is directly affected by the length of spin chains and spin noise, $E_m$ can reach 0.4 at length \emph{M} = 11 and $T_2$ = 1 ms, we find the $E_m$ exhibit exponential dependence on \emph{M} and $T_2$. Unfortunately, because of the numerical limitations, verifying this directly seems to be unrealistic. We have also numerically studied the effect of disorder and spin loss. It is shown that coupling-strength disorder does not have severe impact on $E_m$ when chain is not long (\emph{M} $<$ 11), and chains with odd number are more robust against effect of spin loss than those with even number. Thus it is hopeful for spins star to work under ambient conditions (in the presence of realistic noise and room temperature). Moreover, we have found that star configuration can be applied to quantum detection. The magnitude and orientation of magnetic-field gradient can be both determined by measuring the coherence $\langle C\rangle$ twice.

However, realization of star network is not easy, the imperfection of implantation will definitely cause much disorder, thus the space intervals in spin chain should be designed precisely. Accurate implantation of NV$^-$ and nitrogen impurity require more complicated and precise operation, but it is usually quite difficult for reaching such precision in experiment for now. Another challenge is extension of phase coherence time, actually experimental evidence shows that $T_2$ time of NV$^-$ can be extended to about 2 ms through 128-pulse CPMG control sequences\cite{PhysRevB.86.045214}, $T_2$ time of nitrogen impurity are even shorter\cite{PhysRevLett.101.047601}. Overcoming these difficulties will offer huge possibility to realize star network of spins in solid system.
\section{Acknowledgement.}
We thank Ren-Bao Liu and Erik M. Gauger for their useful discussions.
\bibliographystyle{unsrt}
\bibliography{references}
\end{document}